\documentclass[journal,twocolumn]{IEEEtran}

\usepackage{amsmath,amsfonts,comment}
\usepackage[noadjust]{cite}
\usepackage[pdftex]{graphicx}
\newtheorem{theorem}{Theorem}
\newtheorem{lemma}{Lemma}
\newtheorem{corollary}{Corollary}

\newcommand{\ee}{\mathrm{e}}
\newcommand{\zero}{\mathtt{0}}
\newcommand{\one }{\mathtt{1}}
\newcommand{\zo}{\{\zero,\one\}}
\newcommand{\PP}{\mathbb{P}}
\newcommand{\EE}{\mathbb{E}}
\renewcommand{\vec}[1]{\mathbf{#1}}
\newcommand{\K}{\mathcal K}

\newcommand{\KSSS}{\hat{\mathcal K}_{\mathrm{SSS}}}
\newcommand{\KCOMP}{\hat{\mathcal K}_{\mathrm{COMP}}}
\renewcommand{\L}{\mathcal{L}}

\begin{document}

\title{The Capacity of Bernoulli\\ Nonadaptive Group Testing}
\author{Matthew Aldridge
\thanks{M.~Aldridge is with the Department of Mathematics, University of Bath, Bath, U.K.~and the Heilbronn Institute for Mathematical Research, Bristol, U.K. E-mail: m.aldridge@bath.ac.uk.

Copyright (c) 2017 IEEE. Personal use of this material is permitted.  However, permission to use this material for any other purposes must be obtained from the IEEE by sending a request to pubs-permissions@ieee.org.}}

\maketitle

\begin{abstract}
We consider nonadaptive group testing with Bernoulli tests, where each item is placed in each test independently with some fixed probability. We give a tight threshold on the maximum number of tests required to find the defective set under optimal Bernoulli testing. Achievability is given by a result of Scarlett and Cevher; here we give a converse bound showing that this result is best possible. Our new converse requires three parts: a typicality bound generalising the trivial counting bound, a converse on the COMP algorithm of Chan et al, and a bound on the SSS algorithm similar to that given by Aldridge, Baldassini, and Johnson. Our result has a number of important corollaries, in particular that, in denser cases, Bernoulli nonadaptive group testing is strictly worse than the best adaptive strategies.
\end{abstract}


\section{Main result}


In this paper, we consider nonadaptive noiseless group testing with $n$ items, of which $k$ are defective.
That is, we have $n$ items $\{1, 2, \dots, n\}$, of which a subset $\K$ of size $|\K| = k$ are `defective'. (We assume $k$ is known throughout.) We then try to identify $\K$ through $T$ tests of one or more items: a test is positive if at least one of those items is defective (that is, in $\K$), and negative if none of the items are defective.

Here we consider only Bernoulli nonadaptive testing, where for some fixed $p \in [0,1]$, each item is placed in each test independently with probability $p$. The Bernoulli case has been considered by many authors (\cite{aldridge,aldridge2,atia,chan,malyutov,scarlett,sebo} are just a few examples), as it is simpler to use and to study than some complicated combinatorial designs, and appears to perform well in practice.

Our main result is the following:

\begin{theorem} \label{main1}
Consider nonadaptive noiseless group testing with a Bernoulli test design, and let
  \[ T^* = T^*(n) = \min_{\nu > 0} \max \left\{ \frac{1}{\nu\mathrm{e}^{-\nu}} k \ln k, \frac{1}{h(\mathrm e^{-\nu})} \log_2 \binom nk \right\} \]
  Then, for any $\delta > 0$, with $T > (1+\delta) T^*$ tests, there exists a Bernoulli test design and detection algorithm with error probability tending to $0$ as $n \to \infty$. But with $T < (1 - \delta) T^*$ tests, any Bernoulli test design and detection algorithm will have error probability bounded away from $0$ for sufficiently large~$n$.
\end{theorem}

Here and elsewhere,
\[ h(x) = x \log_2 \frac 1x + (1-x) \log_2 \frac{1}{1-x} \]
is the binary entropy of $x \in [0,1]$. From now on, $\log$ always denotes log to the base $2$, while natural logarithms are explicitly written as $\ln$.

A convenient way to understand the main result (following Baldassini, Johnson and Aldridge \cite{baldassini}) is by looking at the rate $R = \log \binom nk / T$ when $k = \Theta(n^\theta)$ for some sparsity parameter $\theta \in [0,1)$. (Recall that $f(n) = \Theta(g(n))$ if there exists constants $A, B > 0$ such that, for $n$ sufficiently large, $Ag(n) \leq f(n) \leq Bg(n)$.) So, for given $\theta$, a scheme taking $T = c \log \binom nk$
tests, for some constant $c$, has rate $R = 1/c$. One can think of $R$ as the number of bits learned per test, since $\log \binom nk$ is the number of bits required to specify the defective set $\K$. 

A trivial counting bound (see, for example, \cite{baldassini} for a more precise version) shows we require $T \geq \log \binom nk$ tests to find the defective set, which corresponds to the bound $R \leq 1$; that is, we can only learn at most $1$ bit of information from each positive/negative test.

For a given set-up -- in this paper, Bernoulli nonadaptive testing with $k = n^\theta$ defectives for some fixed $\theta \in [0,1)$ -- we say a rate $R$ is achievable if there exists a scheme that finds the defective set $\K$ with $T \leq \log \binom nk / R$ tests with the error probability tending to $0$ as $n \to \infty$. The maximum achievable rate is called the capacity.

Theorem \ref{main1} can then be rewritten as follows:

\begin{theorem} \label{capthm}
For nonadaptive noiseless group testing with $k = \Theta(n^\theta)$ defectives, $\theta \in [0,1)$, and a Bernoulli test design, the capacity is given by
  \begin{equation} \label{cap}
    C(\theta) = \max_{\nu > 0} \min \left\{ \frac{\nu\mathrm{e}^{-\nu}}{\ln 2} \frac{1-\theta}{\theta}, h(\mathrm e^{-\nu}) \right\} .
  \end{equation}
\end{theorem}

\begin{figure} 
\begin{center}
\includegraphics[width=0.475\textwidth]{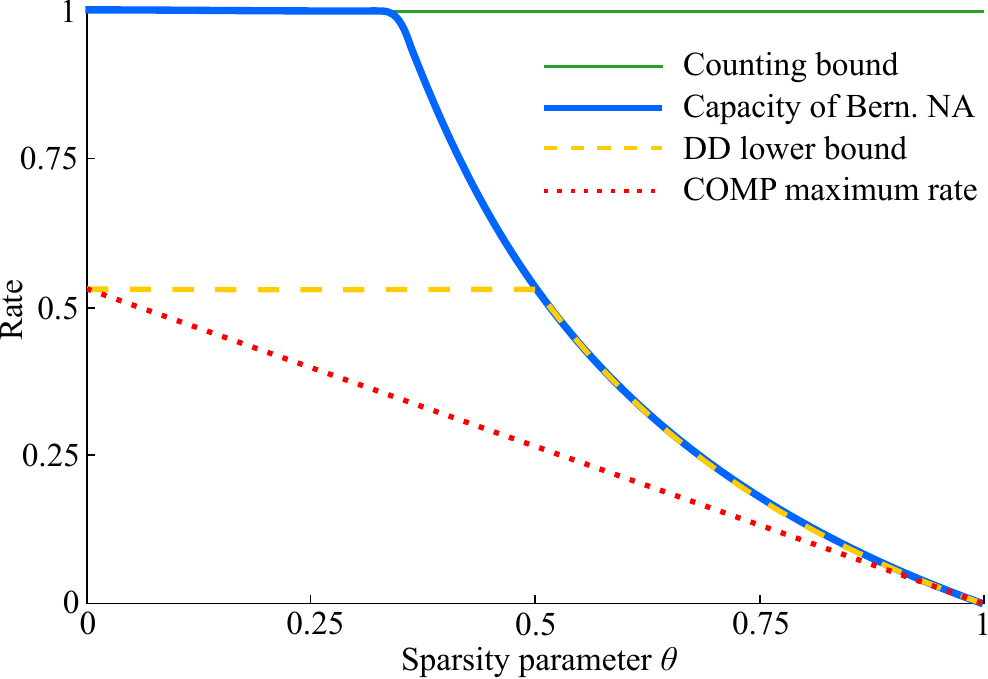}
\end{center}
\caption{Graph showing group testing rates with $k = \Theta(n^\theta)$ defectives: the trivial counting bound $C \leq 1$; the capacity of Bernoulli nonadaptive group testing (Theorem \ref{capthm}); an achievable rate of the DD algortihm \eqref{ddrate} \cite{aldridge}; and the maximum achievable rate of the COMP algorithm with Bernoulli tests (Corollary \ref{compcor})} \label{mainfig}
\end{figure}

It's not difficult to check that for $\theta \leq 1/3$, the outer maximum is achieved at $\nu = \ln 2$, leaving the second minimand to dominate, to get $C = 1$.   For $\theta \geq \theta^*$, where
  \[ \theta^* = \frac{1}{1 + h(\mathrm e^{-1})\mathrm e \ln 2} \approx 0.359 \]
the the outer maximum is achieved at $\nu = 1$, with the first minimand in \eqref{cap} dominating, to get
  \[ C = \frac{1}{\mathrm e \ln 2} \frac{1-\theta}{\theta} \approx 0.531 \frac{1-\theta}{\theta} . \]
Hence, it's only in the narrow range $\theta \in (1/3, \theta^*)$ that the complicated maximin is necessary. The capacity \eqref{cap} and other important rates are shown in Figure \ref{mainfig}.

In \eqref{cap}, $\nu$ is a parameter representing the Bernoulli parameter $p$ through the relationship $p = 1 - \ee^{-\nu/k}$, so $p$ scales like $\nu / k$ for $\theta > 0$. Note that $\nu = \ln 2$ gives $p = 1 - 2^{-1/k}$, so $(1 - p)^k = 1/2$, meaning tests are $50:50$ positive or negative, while $\nu = 1$ gives $p$ scaling like $1/k$, meaning tests contain an average of one defective item.


The capacity \eqref{cap} in Theorem \ref{capthm} is equal to an achievable rate due to Scarlett and Cevher \cite[Corollary 6]{scarlett}. (In \cite{scarlett2}, Scarlett and Cevher emphasise the group testing aspects of their wide-ranging results from \cite{scarlett}.) The new work in this paper is to give the converse result, showing that this cannot be improved. As far as we are aware, this is the first group testing converse of any sort that gives tighter rate bounds than that given by the counting bound (or its equivalents for group testing with noise). Scarlett and Cevher note \cite{scarlett2}, combining their achievability result with the trivial counting bound $C \leq 1$ suffices to establish Theorem \ref{main1} for $\theta \leq 1/3$; the new work in this paper is to prove it for $\theta > 1/3$.

Our converse requires three parts:
\begin{enumerate}
  \item A typicality bound, which generalises the trivial counting bound $C \leq 1$ (Subsection \ref{typbd});
  \item A new converse on the COMP algorithm of Chan, Che, Jaggi and Saligrama \cite{chan} (Subsection \ref{compbd}); 
  \item A converse on the SSS algorithm very similar to that of Aldridge, Baldassini and Johnson \cite{aldridge} (Subsection \ref{sssbd});
\end{enumerate}
together with an argument that these three points suffice to prove the desired converse result (Subsection \ref{outline}).

Further, our result (and its proof) carries a number of important corollaries:
\begin{itemize}
  \item For $\theta > 1/3$, Bernoulli nonadaptive testing is strictly worse than the best adaptive schemes. This in contrast to the fact that, as Scarlett and Cevher note \cite{scarlett2}, Bernoulli nonadaptive group testing achieves the same rate $C=1$ as the best adaptive schemes for $\theta \leq 1/3$. 
  \item The maximal rate achievable by the COMP algorithm of Chan et al is $(1 - \theta)/\ee\ln 2 \approx 0.531 (1 - \theta)$. In particular, COMP is a suboptimal detection algorithm for Bernoulli nonadaptive  group testing for all $\theta \in [0,1)$.
  \item The SSS algorithm (while probably impractical) is an optimal detection algorithm for Bernoulli nonadaptive group testing for all $\theta$.
  \item The practical DD algorithm of Aldridge et al \cite{aldridge} is an optimal detection algorithm for Bernoulli nonadaptive  group testing for $\theta \in [1/2,1)$.
\end{itemize}
We discuss these corollaries further in Section \ref{corrs}.

\section{Notation and algorithms}

In this section, we fix some notation (which readers familiar with group testing may wish to skip). We then discuss the COMP and SSS algorithms, which will be important in the proof of our main result later.

\subsection{Group testing notation}

Recall that we have $n$ \emph{items}, of which a set $\K\subseteq \{1,2,\dots, n\}$ of size $k = |\K|$ are \emph{defective}. We test them with $T$ \emph{tests}. We summarise these tests through a \emph{test design} $\mathsf X = (x_{it}) \in \zo^{n \times T}$, where $x_{it} = \one$ denotes that item $i$ is in test $t$, and $x_{it} = \zero$ denotes that it is not. Upon testing, we receive back the \emph{outcomes} $\vec y = (y_t) \in \zo^T$ according to the rule
  \begin{equation} \label{gtrule}
    y_t = \begin{cases} \one &  \text{if for some $i \in \K$ we have $x_{it} = \one$,} \\
                        \zero & \text{if for all $i \in \K$ we have $x_{it} = \zero$,} \end{cases}
  \end{equation} 
where $y_t = \one$ denotes a positive test, and $y_t = \zero$ denotes a negative test.

In this paper we consider \emph{Bernoulli nonadaptive group testing}. This means that, for some fixed $p \in [0, 1]$ we have $x_{it} = \one$ with probability $p$ and $x_{it} = \zero$ with probability $1 - p$, independently over $i$ and $t$.

Then given the testing matrix $\mathsf X$ and the outcomes $\vec y$ (including knowledge of the number of defectives $k$ but not the defective set $\K$), we must estimate the true defective set. A \emph{detection algorithm} is a map $\hat{\K} \colon \zo^T \to \mathcal P \{1,2,\dots, n\}$, where $\mathcal P$ denotes the power-set (the set of all subsets). 

The \emph{average error probability} of the detection algorithm $\hat{\K}$ for Bernoulli$(p)$ nonadaptive group testing is
  \[ \epsilon = \frac{1}{\binom nk} \sum_{\K : |K| = k} \PP ( \hat{\K}(\vec y) \neq \K \mid \text{defective set $\K$} ) , \]
where the probability is over the the Bernoulli$(p)$ testing matrix $\mathsf X$, and $\vec y$ is related to $\mathsf X$ and $\K$ through \eqref{gtrule}. The \emph{success probability} is $1 - \epsilon$.

In this paper we consider a sequence of group testing procedures, indexed by the number of items $n$, going to $\infty$. We consider the case where $k = k(n)$ grows like $o(n)$, and in rate calculations, more specifically when $k = \Theta(n^\theta)$ for $\theta \in [0, 1)$. Note that our result (Theorem \ref{main1}) is already proven for $k$ bounded (the $\theta = 0$ case), by combining the trivial counting bound with the achievability result at $\theta = 0$ (originally by Malyutov \cite{malyutov2} and Seb\H o \cite{sebo}, but also as special cases of the results of Aldridge et al \cite{aldridge2} or Scarlett and Cevher \cite{scarlett,scarlett2}).  Thus we are safe to consider $k \to \infty$ (that is, $\theta > 0$) where convenient. We will use $T = T(n)$ tests, and are looking for whether the error probability $\epsilon = \epsilon(n)$ tends to $0$ or is bounded away from $0$.

Recall that the \emph{rate} of a group testing procedure is defined as $\binom nk / T$, and that a rate $R$ is achievable if there exists a sequence of procedures with rate at least $R$ with $\epsilon \to 0$. We refer to the maximum rate by Bernoulli nonadaptive group testing as the \emph{capacity}.

Given a test design $\mathsf X$ and outcomes $\mathbf y$, we call a set $\L \subseteq \{1,2,\dots,n\}$ \emph{satisfying} if no negative test contains an item from $\L$ and every positive test contains at least one item from $\L$. In other words, we would have got the same output $\vec y$ had $\L$ been the true defective set.

Clearly the defective set $\K$ itself is a satisfying set, but there can be other satisfying sets too. Given the test design and outcomes, the posterior distribution of the defective set is uniform over all satisfying sets of the correct size $k$. Hence accurate detection of the defective set requires there to usually be only one satisfying set of size $k$.

\medskip

Recall we are writing $\log$ for logarithms to base $2$. We write $\simeq$ for asymptotic equivalence; that is, $f(n) \simeq g(n)$ if $f(n) = \big(1 + o(1)\big)g(n)$. A useful asymptotic expression for $\theta < 1$ will be
  \[ \log \binom nk \simeq k \log \frac nk . \]

\subsection{The COMP algorithm}

The COMP algorithm is based on the following observation: any item that appears in a negative test is certainly nondefective. The COMP algorithm therefore declares all such items to be nondefective, and declares all the remaining items to be defective.\footnote{We write `COMP' to refer to the detection algorithm given in \eqref{comp}, which can be used with any test design $\mathsf X$. We use the name `COMP' following Chan et al \cite{chan}, although they use the term to mean the detection algorithm \eqref{comp} \emph{only} when twinned with a Bernoulli nonadaptive test design. They use the term `CBP' to denote the same detection algorithm \eqref{comp} -- but explained differently -- when used with a slightly different test design. In a later paper \cite{chan2}, a similar set of authors refer to COMP and CBP respectively by the different names `CoMa' and `CoCo'. Chan et al \cite{chan2} note that variants on COMP have appeared many times in the literature under various names, for example, \cite{malyutov,chen,kautz}.}  That is, COMP declares
  \begin{equation} \label{comp}
    \KCOMP = \{1, \dots, n \} \setminus \{ i : \exists\, t \text{ such that } x_{it} = \one \text{ and } y_t = \zero \}
  \end{equation}
as its guess of the defective set $\K$.

The following three facts about COMP are immediate and will be important:
\begin{itemize}
  \item $\KCOMP \supseteq \K$, since no actual defectives can have been ruled out;
  \item $\KCOMP$ is a satisfying set, since by definition no items of $\KCOMP$ are in a negative test, and the true defectives in $\K$ ensure an item from $\KCOMP$ in each positive test;
  \item $\KCOMP$ is the unique largest satisfying set, since all items not in $\KCOMP$ are ruled out through their appearance in negative tests.
\end{itemize}

Chan et al \cite{chan} show that COMP can succeed with $T = (1+\delta) \ee k \ln n$ tests, for an achievable rate of
  \[ \frac{1}{\ee \ln 2} (1 - \theta) \approx 0.531 (1 - \theta) . \]
(Aldridge et al \cite[Remark 18]{aldridge} give an alternative proof.) We show later (Lemma \ref{complem}) that this is in fact the maximum rate achievable by COMP.

\subsection{The SSS algorithm}

The SSS algorithm chooses the smallest satisfying set as its guess of $\K$. (For the purposes of this paper, it is not important what the algorithm does if the smallest satisfying set is not unique -- it suffices to note that the error probability in this situation is at least $1/2$.) The SSS algorithm was studied in some detail by Aldridge et al \cite{aldridge}.

Note that although we refer to SSS as an `algorithm', we do not claim that finding the smallest satisfying set is computationally feasible for large problems. Aldridge et al \cite[Subsection III.3]{aldridge} discuss the connections between SSS and the NP-complete set-cover problem, and propose an algorithm they call SCOMP as a `best practical approximation' to SSS.

\section{Proof of main result}

We now prove our main theorem (Theorem \ref{main1}).

\subsection{Outline of proof} \label{outline}

The achievability was proved by Scarlett and Cevher \cite[Corollary 6]{scarlett} \cite{scarlett2}. (The result was previously known in the special cases $\theta = 0$ \cite{malyutov2,sebo,aldridge2}, and for $\theta \in [1/2,1)$ using the DD algorithm \cite{aldridge}.) We are required to prove a converse. 

The crucial idea is the following. First, if the (necessarily unique) largest satisfying set is of size $k$, then it must be the true defective set $\K$, and it is found by the COMP algorithm. Second, if the smallest satisfying set is of size $k$ and is unique (which it might or might not be), then it is the true defective set $\K$, and it is found by the SSS algorithm. Third, if neither of these two cases occurs, there is more than one satisfying set of size $k$, and our error probability is at least $1/2$. Hence, if we can show that both the SSS algorithm and the COMP algorithm fail above a certain rate, then that rate is not achievable, and we have a converse.

Let us now be more formal. Recall that the SSS algorithm chooses the smallest set of items $\KSSS$ that satisfies the test outputs. Recall also (as noted above and in \cite{aldridge}) that the COMP algorithm chooses the largest set of items $\KCOMP$ that satisfies the test outputs. Further any set $\mathcal L$ satisfying $\KSSS \subseteq \mathcal L \subseteq \KCOMP$ is also a satisfying set.
Hence we also have the following: if both $|\KSSS| < k$ and $|\KCOMP| > k$ (noting the strict inequality), then there is not a unique satisfying set of size $k$, since there are at least two such $\L$s. In this situation, our average probability of error must be at least $1/2$, since we can do no better than arbitrarily choose one of the (at least two) satisfying sets of the correct size $k$.

Hence, if we can show for rates above $C(\theta)$ in \eqref{cap} that
  \[ \mathbb P ( |\KSSS| < k \text{ and } |\KCOMP| > k) \]
is bounded away from $0$, then we are done. 

We can write
  \begin{align}
              \mathbb P ( |\KSSS| &< k   \text{ and }  |\KCOMP| > k) \notag \\
    &\quad =    1 - \mathbb P ( |\KSSS| = k   \text{ or  }  |\KCOMP| = k) \notag \\
    &\quad \geq 1 - \mathbb P ( |\KSSS| = k ) - \mathbb P ( |\KCOMP| = k) , \label{idea}
  \end{align}
where we have used the union bound.

Note that $\mathbb P ( |\KCOMP| = k )$ is precisely the probability that COMP succeeds in finding the defective set, as $\KCOMP$ will be the only satisfying set of the correct size, since it is the unique largest satisfying set. We can also bound $\PP (|\KSSS| = k)$ by looking at a particular event that causes SSS to fail: specifically that some $\mathcal J \subset \K$ of size $|\mathcal J| = k-1$ is also satisfying, due to one of the defectives never being the only defective item in a test.

What remains is to prove that $\mathbb P ( |\KCOMP| = k )$ and $\mathbb P ( |\KSSS| = k )$ are both small enough. We prove that $\mathbb P ( |\KCOMP| = k ) \leq 3/8$ in Lemma \ref{complem}, and prove that $\mathbb P ( |\KSSS| = k ) \leq 1/2$ in Lemma \ref{ssslem}. This completes the proof. \hfill $\IEEEQED$

\subsection{Typicality converse} \label{typbd}

The following typicality bound generalises the trivial counting bound that one requires $T > \log_2 \binom nk$. In particular, this lemma is necessary to prove Lemma \ref{ssslem} later.

\begin{lemma} \label{lemtyp}
Consider nonadaptive noiseless group testing with a Bernoulli$(p)$ test design. Write
  \[ T_{\text{typ}} \geq 
       \frac{1}{h\big( (1-p)^k \big)} \log \binom nk . \]
Then with $T \leq (1 - \delta) T_{\text{typ}}$ tests, the error probability is bounded away from $0$.
\end{lemma}

The idea is the following. The trivial counting bound comes from noting that the number of defective sets $\binom nk$ must be no larger than the number of possible outcomes $2^T$. However, the theory of typical sets tells us that in fact we can only see about $2^{Th((1-p)^k)}$ outcomes, since $(1 - p)^k$ is the probability of a negative outcome. Hence we require $2^{Th((1-p)^k)} \geq \binom nk$, and the result follows.

Or, for alternative way of thinking about it: the counting bound tells us we can only learn one bit from each positive-or-negative test. But this is only if the outcome is $50:50$; with probability $(1 - p)^k$ of a negative test, we can only learn $h((1 - p)^k)$ bits per test.

Note that setting $p = 1 - 2^{-1/k}$ in Lemma \ref{lemtyp}, so that $(1 - p)^k = 1/2$, and recalling that $h(1/2) = 1$, we recover the trivial counting bound as a special case.

We now formalise this intuition.

\begin{IEEEproof}
Consider a Bernoulli$(p)$ test design where there are $k$ defectives. Then clearly a test is negative if none of the $k$ defectives appear, with probability $\pi = (1 - p)^k$, and is positive otherwise, with probability $1 - \pi$. In what follows a capital $\vec Y$ denotes a random vector of test outcomes (over the random defective set and test design), while a lower case $\vec y$ will denote a particular value in $\zo^T$ that $\vec Y$ could take.

Write $\mathcal T(\delta)$ for the set of $\delta$-typical outputs $\vec y$; that is,
  \begin{align*}
    \tilde H(\vec y) &= \frac{\#\{t : y_t = \zero\}}T \log_2 \frac{1}{\pi} + \frac{\#\{t : y_t = \one \}}T \log_2 \frac{1}{1 - \pi} , \\
    \mathcal T(\delta) &= \left\{ \vec y \in \zo^T : \big|\tilde H(\vec y) - h(\pi)\big| \leq \delta \right\}.
  \end{align*}
Then standard results on typical sets (see for example \cite[Chapter 3]{cover}) give that
\begin{enumerate}
  \item $\PP (\vec Y \not\in \mathcal T(\delta)) \to 0$ as $T \to \infty$,
  \item $|\mathcal T(\delta)| \leq 2^{T(h(\pi)+\delta)}$
\end{enumerate}

We can write the success probability as
  \begin{align*}
    \PP(\text{success})
      &= \PP (\vec Y     \in \mathcal T(\delta)) \PP(\text{success} \mid \vec Y     \in \mathcal T(\delta)) \\
                     &\qquad {}+ \PP (\vec Y \not\in \mathcal T(\delta)) \PP(\text{success} \mid \vec Y \not\in \mathcal T(\delta)) \\
      &\leq \PP(\text{success} \mid \vec Y \in \mathcal T(\delta)) + \PP (\vec Y \not\in \mathcal T(\delta)) .
  \end{align*}
From point 1.\ above, we know that the second term is small for $T$ sufficiently large. We claim that the first term will be bounded away from $1$ unless
  \[ T > (1 - \delta) \frac{1}{h(\pi)} \log_2 \binom nk . \]
A standard bound on the binomial coefficient completes the proof.

We now prove the claim, in a manner similar to Baldassini et al's proof of a `strong' version of the trivial counting bound \cite[Theorem 3.1]{baldassini}.

Fix a test design for $n$ items in $T$ tests. Then, given some $\L \subseteq \{1,2,\dots, n\}$ with $|\L| = k$, write $A(\L) \in \zo^T$ for vector of test outcomes given the test design and were the defective set $\L$. Further, for $\vec y \in \zo^T$, write
  \[ \mathcal S(\vec y) = A^{-1}(\vec y) = \{ \L : |\L| = k \text{ and }A(\L) = \vec y \} \]
for the collection of satisfying sets of size $k$, and $s(\vec y) = |\mathcal S(\vec y)|$ for the number of satisfying sets. Note that if $\vec y$ is the outcome of the group testing procedure, we cannot have $s(\vec y) = 0$, and otherwise the probability of success is at most $1/s(\vec y)$. Then, using $\mathbb I[\ ]$ for the indicator function, the probability of success is bounded by
  \begin{align*}
    &\PP(\text{success} \mid \vec Y \in \mathcal T(\delta)) \\
      &\quad= \sum_{|\L| = k} \frac{1}{\binom nk} \PP(\text{success} \mid \text{defective set } \L, \vec Y \in \mathcal T(\delta) ) \\
      &\quad= \frac{1}{\binom nk} \sum_{|\L| = k} \sum_{\vec y \in \mathcal T(\delta)} \mathbb I[A(\L) = \vec y] 
      \PP(\text{success} \mid \text{def. set } \L ) \\
      &\quad\leq \frac{1}{\binom nk} \sum_{|\L| = k} \sum_{\vec y \in \mathcal T(\delta)} \mathbb I[A(\L) = \vec y] \frac{1}{s(\vec y)} \\
      &\quad= \frac{1}{\binom nk} \sum_{\vec y \in \mathcal T(\delta)} \frac{1}{s(\vec y)} \sum_{|\L| = k} \mathbb I[A(\L) = \vec y] \\
      &\quad= \frac{1}{\binom nk} \sum_{\vec y \in \mathcal T(\delta)} \frac{1}{s(\vec y)} s(\vec y) \\
      &\quad= \frac{|\mathcal T(\delta)|}{\binom nk} .
  \end{align*}
From point 2.\ above, we have $|\mathcal T(\delta)| \leq 2^{T(h(\pi)+\delta)}$. Substituting this in and rearranging (including resetting $\delta$) gives the claim, and we are done.
\end{IEEEproof}

Note that, in particular, any Bernoulli nonadaptive algorithm that achieves a nonzero rate must, therefore, use $p = p(n)$ such that $(1 - p)^k$ is bounded away from $0$ and from $1$. In particular, this means we are safe to assume in what follows that we always have $p = \Theta(1/k)$.

\subsection{Converse to COMP} \label{compbd}

Recall that the COMP algorithm of Chan et al \cite{chan} works as follows: any item in a negative test is definitely nondefective, and we declare all other items to have been defective. Hence, for COMP to succeed, it is necessary for every nondefective to appear in a negative test.

We have the following converse for the COMP algorithm

\begin{lemma} \label{complem}
Consider noiseless nonadaptive Bernoulli group testing, decoding using the COMP algorithm. Write
\[ T^*_{\mathrm{COMP}} = T^*_{\mathrm{COMP}}(n) = \mathrm ek \ln n . \]
Then for all $\delta$, with $T < (1 - \delta)T^*_{\mathrm{COMP}}$ tests, for any $p = p(n)$, Bernoulli nonadaptive testing using the COMP algorithm has error probability at least $5/8$.
\end{lemma}

Note that $T > (1+\delta)T^*_{\mathrm{COMP}}$ was shown to be achievable by Chan et al \cite{chan} (with Aldridge et al \cite[Remark 18]{aldridge} providing an alternative proof), so this is a tight threshold. 

\begin{IEEEproof}
For COMP to correctly find the defective set, we require each of the $n - k$ nondefectives to appear in a negative test. We seek to bound the probability that this occurs.

Let us write $q = 1-p$ for brevity. The number $M$ of negative tests is $M \sim \text{Bin}(T, q^k)$, as each of the $T$ tests is negative if and only if all $k$ defectives fail to appear in it. Then, conditional on $M$, the probability that a nondefective item appears in at least one of those tests is $1 - q^M$, and the probability that all $n-k$ do is $(1 - q^M)^{n-k}$. Hence we have
  \[
    \mathbb P (\text{success})
      = \mathbb E (1 - q^M)^{n-k} 
      \leq \mathbb E (1 - q^M)^{n'} .
  \]
Here we are writing $n' = (1 - \delta)n$, and using the fact that, since $k = o(n)$, for $n$ sufficiently large we have $n - k \geq n'$.

For any $m$ we have
  \begin{align}
    \mathbb P (\text{success})
      &\leq \mathbb E (1 - q^M)^{n'} \notag \\
      &= \PP(M \leq m) \EE \big( (1 - q^M)^{n'} \mid M \leq m \big) \notag \\
           &\qquad {}+ \PP(M > m) \EE \big( (1 - q^M)^{n'} \mid M > m \big) \notag \\
      &\leq \EE \big( (1 - q^M)^{n'} \mid M \leq m \big) + \PP(M > m) \notag \\
      &\leq (1 - q^m)^{n'} + \PP(M > m) , \label{comp1}
  \end{align}
where we are using that $(1 - q^m)^{n'} \leq 1$ and is increasing in $m$. By choosing $m = (1+\delta)\EE M = (1+\delta)Tq^k$, standard concentration of measure results (the Azuma--Hoeffding inequality, for example) the term $\PP(M > m)$ can be made arbitrarily small as $n \to \infty$. It remains to bound the first term in \eqref{comp1}.

Expanding this out to five terms,\footnote{It would be natural to expand out to three terms to get an upper bound here. However, this leads to a bound on the success probability of $1/2$. Combined with a similar bound of $1/2$ for the SSS algorithm later, this would insufficient for a nontrivial bound in \eqref{idea}. Hence, we expand to five terms here, to get a slightly tighter bound.} we get the bound
  \begin{align}
    & (1 - q^m)^{n'} \\
      &\quad \leq 1 - n'q^m + \binom{n'}{2} q^{2m} - \binom{n'}{3} q^{3m}
              + \binom{n'}{4} q^{4m} \notag \\
      &\quad= 1 - n'q^m \bigg(1 - \frac{1}{2} (n'-1)q^{m} \bigg(1 - \frac{1}{3} (n'-2)q^{m} \notag \\
      & \qquad\qquad\qquad\qquad\qquad\quad\ \times  \bigg(1 - \frac{1}{4} (n'-3)q^{m} \bigg) \bigg) \bigg) \label{comp3}
  \end{align}

Writing $p = 1 - \ee^{-\nu/k}$, so $q^k = \ee^{-\nu}$, and recalling we had set $m = 
(1+\delta)Tq^k$, we have that
  \[
    n' q^m = \exp \left( \ln n + \ln(1 - \delta) - (1+\delta) \frac{\nu \ee^{-\nu}}{k} T \right) .
  \]
Noting that the error probability decreases with $T$, we see that~for
  \begin{equation}
    T \leq (1 - \delta) \frac{1}{\nu \ee^{-\nu}}k \ln n \label{comp4}
  \end{equation}
we have from \eqref{comp3} that
  \begin{align*}
  (1 - q^m)^{n'} &\leq  1 - \left(1 - \frac{1}{2} \left(1 - \frac{1}{3} \left(1 - \frac{1}{4} \right) \right) \right) + \eta \\
  &= \frac38 + \eta
  \end{align*}
for $\eta$ arbitrarily small and $n$ sufficiently large. This can be substituted into  \eqref{comp1}.

Finally we note that the bound \eqref{comp4} is optimised at $\nu = 1$, giving the desired result.
\end{IEEEproof}

\subsection{Converse to SSS} \label{sssbd}

\begin{lemma} \label{ssslem}
Consider noiseless nonadaptive Bernoulli group testing, decoding using the SSS algorithm. Write
  \begin{equation} \label{sssbdd}
    T^* = \min_{\nu > 0} \max \left\{ \frac{1}{\nu\mathrm{e}^{-\nu}} k \ln k, \frac{1}{h(\mathrm e^{-\nu})} k \log_2 \frac{n}{k} \right\} .
  \end{equation}
Then for all $\delta$, with $T < (1 - \delta)T^*$ tests, for any $p = p(n)$, Bernoulli nonadaptive testing using the SSS algorithm has error probability at least $1/2$.
\end{lemma}

\begin{IEEEproof}
The second term in the minimum in \eqref{sssbdd} is the typicality bound from Lemma \ref{lemtyp} under the substitution $p = 1 - \ee^{-\nu/k}$, so that $(1 - p)^k = \ee^{-\nu}$. It remains to show the bound in the first term. Optimising over $\nu$ then gives the result.

For the first term in \eqref{sssbdd}, we follow the argument of \cite[Lemma 28]{aldridge}.\footnote{In fact the required result virtually appears in \cite{aldridge}: if one follows the arguments towards Lemma 28 of that paper, using its Lemmas 24, 30, 31 on the way, but declines the simplifications and bounds it gains by specifying $p = 1/k$ early in the argument, one gets the result required here. (Note also a small typographical error in \cite{aldridge}, where its applications of the inclusion--exclusion formula in its equations (47) and (48) are missing a $(-1)^\ell$ term.) For completeness, we give here the full argument, but somewhat briefly -- the reader desiring complete details is directed to \cite{aldridge}.}

The key point is the following: the SSS algorithm will only succeed if for each defective item there is a (necessarily positive) test in which it is the \emph{only} defective item -- otherwise, the item could be removed to give a smaller satisfying set. Note that if this doesn't happen -- if there is some defective item that only ever appears with other defective items -- then we have $\KSSS \leq k-1$, so we are bounding $\mathbb P(|\KSSS| \neq k)$ here, as required for the larger argument.

If we write $A_i$ for the event that defective $i \in \K$ is the sole defective in some test, we have
  \begin{align*}
    \PP(\text{SSS succeeds})
      &\leq  \PP \left( \bigcap_{i \in \K} A_i \right) \\
      &= 1 - \PP \left( \bigcup_{i \in \K} A_i^C \right) \\
      &= 1 - \sum_{j=1}^k (-1)^{j+1} \sum_{|\mathcal J| = j} \PP \left( \bigcap_{i \in \mathcal J} A_i^C \right) ,
  \end{align*}
by the inclusion--exclusion formula. Thus we have
  \begin{align*}
    \PP(\text{error })
      &\geq \sum_{j=1}^k (-1)^{j+1} \sum_{|\mathcal J| = j} \PP \left( \bigcap_{i \in \mathcal J} A_i^C \right) \\
      &= \sum_{j=1}^k (-1)^{j+1} \binom kj (1 - j p(1-p)^{k-1} )^T ,
  \end{align*}
where we have used that $\bigcap_{i \in \mathcal J} A_i^C$ is precisely the event that each of the $T$ tests fails to include a unique item from $\mathcal J$, which has probability $1 - j p(1-p)^{k-1}$.

Bounding the same way as we did in the proof of Lemma \ref{complem}, this time expanding out only two terms, gives
  \begin{align*}
    \PP(\text{error})
      &\geq \sum_{j=1}^k (-1)^{j+1} \binom kj (1 - j p(1-p)^{k-1} )^T \\
      &= \sum_{j=1}^k (-1)^{j+1} \binom kj (1 - j r )^T \\
      &\geq k(1 - r)^T - \frac12 k^2 (1 - 2r)^T \\
      &= k(1 - r)^T \left( 1 - \frac12 k \left( \frac{1 - 2r}{1 - r} \right)^T \right)
  \end{align*}
where we are writing $r = p(1-p)^{k-1}$ for brevity. We now use the bound $(1 - x)^T \leq \ee^{-xT}$ for $x \leq 1$ in two ways. First with $x = -r/(1-r)$ (which is negative) and second with $x = r/(1 - r)$ (which tends to $0$, so is eventually less than $1$), to get
  \begin{align*}
    &\PP(\text{error}) \\
      &\ \geq k \exp\left( -\frac{r}{1-r}T\right) \left( 1 - \frac12 k \exp\left(-\frac{r}{1-r}T\right) \right) \\
      &\ = \exp\left(\ln k -\frac{r}{1-r}T\right) \left( 1 - \frac12 \exp\left(\ln k -\frac{r}{1-r}T\right) \right) ,
  \end{align*}
for $n$ sufficiently large.

We thus see that for 
  \[ T \leq \frac{1 - r}{r} \ln k + 1 \]
we have
  \[ \PP(\text{error}) \geq \exp\left( -\frac{r}{1-r}\right) \left( 1 - \frac12 \exp\left(-\frac{r}{1-r}\right) \right) ,\]
since the error probability is nonincreasing as $T$ decreases.
Writing $p = 1 - \ee^{-\nu/k}$, we see that $p \simeq \nu/k$, so $r \simeq \nu \ee^{-\nu} /k$, and hence
  \[ \frac{1-r}{r} \simeq \frac{1}{\nu \ee^{-\nu}} k - 1\simeq \frac{1}{\nu \ee^{-\nu}} k . \]
Thus we see the bound on the error probability tends to $1/2$ for
  \[ T \leq (1-\delta)\frac{1}{\nu \ee^{-\nu}}k \ln k \]
giving the second term in \eqref{sssbdd}.
\end{IEEEproof}

\section{Corollaries and recent work} \label{corrs}

We now briefly explain some corollaries to our work, as listed in the introduction. Many of these are immediate upon examining Figure \ref{mainfig}. We also briefly discuss some more recent work that appeared while this paper was in review.

In what follows, we use `optimal' to mean `achieves the optimal rate', and not in any stronger sense.

\begin{corollary}
  For $\theta \leq 1/3$, Bernoulli nonadaptive testing can achieve the same rate as an optimal adaptive test design, whereas for $\theta > 1/3$, Bernoulli nonadaptive testing cannot achieve the same rate as an optimal adaptive test design.
\end{corollary}

\begin{IEEEproof}
As Scarlett and Cevher note \cite{scarlett2}, for $\theta \leq 1/3$, their achievability result of rate $1$ combined with the trivial counting bound $C \leq 1$ shows that Bernoulli nonadaptive testing is optimal.

For $1/3 < \theta < 1$, it's simple to see from \eqref{cap} that our new converse shows that $C < 1$ for Bernoulli nonadaptive testing. However, as explicitly noted by Baldassini et al \cite{baldassini}, $C = 1$ can be achieved adaptively using Hwang's binary splitting algorithm \cite{hwang} \cite[Section 2.4]{duandhwang}.
\end{IEEEproof}

An important open question here is what the capacity of nonadaptive testing is if we remove the requirement for tests to be Bernoulli. Johnson, Aldridge, and Scarlett \cite{jas} have proved that `constant tests-per-items' designs outperform Bernoulli designs for $\theta > 0.434$, and nonrigorous arguments from statistical physics \cite{mezard} suggest this should be true for all $\theta > 1/3$.

\begin{corollary} \label{compcor}
 The maximal rate achievable by the COMP scheme of Chan et al is $(1 - \theta)/\ee\ln 2 \approx 0.531 (1 - \theta)$. In particular, COMP is a suboptimal detection algorithm for nonadaptive Bernoulli group testing for all $\theta \in [0,1)$.
\end{corollary}

\begin{IEEEproof}
Chan et al \cite[Theorem 4]{chan} prove achievability, and our Lemma \ref{complem} gives the converse. This is clearly suboptimal compared to \eqref{cap}.
\end{IEEEproof}

\begin{corollary}
The SSS algorithm is an optimal detection algorithm for nonadaptive Bernoulli group testing for all $\theta$.
\end{corollary}

\begin{IEEEproof}
Our `proof outline' shows that one can only succeed if either the defective set $\K$ is the largest satisfying set, in which case COMP finds it; or $\K$ is the smallest satisfying set, in which case SSS finds it. However, Corollary \ref{compcor} shows that COMP is not optimal, therefore SSS must be.
\end{IEEEproof}

\begin{corollary}
The DD algorithm of Aldridge et al \cite{aldridge} is an optimal detection algorithm for nonadaptive Bernoulli group testing for $\theta \in [1/2,1]$.
\end{corollary}

\begin{IEEEproof}
  Aldridge et al \cite[Theorem 12]{aldridge} show that the DD algorithm can achieve the rate
  \begin{equation} \label{ddrate}
    \frac{1}{\ee \ln 2} \min \left\{ \frac{1 - \theta}{\theta}, 1\right\} .
  \end{equation}
For $\theta \geq 1/2$, the first term is the minimum, which is the same as the capacity result in \eqref{cap}.
\end{IEEEproof}

An open question is if DD, or any other practical algorithm, achieves the Bernoulli nonadaptive capacity \eqref{cap} for $\theta < 1/2$. Aldridge \cite{aldridgerate} gave an upper bound on the rate of DD showing that it is strictly suboptimal for $\theta < 0.407$.

Another open problem is to produce better converses for Bernoulli nonadaptive group testing with noisy tests. Scarlett and Cevher \cite{scarlett,scarlett2} give similar achievability results, but the only converses known are along the lines of the counting bound. However, the new converses given in this paper relied heavily on the tests being noiseless, so it is not clear if our arguments can be generalised to the noisy case.

\section*{Acknowledgements}

The author thanks Oliver Johnson  and Jonathan Scarlett for useful discussions.
 
\IEEEtriggeratref{6}

\bibliographystyle{IEEEtran}
\bibliography{cap-bib}

\end{document}